%% file: main.tex
\documentclass[conference,letterpaper]{IEEEtran}

\usepackage{geometry}
\usepackage[utf8]{inputenc}
\usepackage[T1]{fontenc}
\usepackage{url}
\usepackage{ifthen}
\usepackage{cite}
\usepackage[cmex10]{amsmath}
\usepackage{amsthm}
\usepackage{tikz}
\usepackage{amssymb}
\usepackage{bm}
\usepackage{graphicx}
\usepackage{booktabs}
\usepackage{algorithm}
\usepackage{algorithmic}
\usepackage{flafter}
\usepackage{enumitem}
\newlength{\savedtextfloatsep}
\newtheorem{lemma}{Lemma}
\newtheorem{proposition}{Proposition}

\usepackage{tikz}
\usetikzlibrary{
  arrows.meta,
  positioning,
  calc,
  fit,
  backgrounds,
  shapes.geometric,
  decorations.pathmorphing,
  decorations.pathreplacing,
  patterns
}
\definecolor{certBlue}{HTML}{1F618D}
\definecolor{certBlueLight}{HTML}{EBF5FB}
\definecolor{certOrange}{HTML}{CA6F1E}
\definecolor{certOrangeLight}{HTML}{FEF5E7}
\definecolor{certGreen}{HTML}{1E8449}
\definecolor{certGreenLight}{HTML}{EAFAF1}
\definecolor{certRed}{HTML}{B03A2E}
\definecolor{certRedLight}{HTML}{FDEDEC}
\definecolor{certGray}{HTML}{566573}
\definecolor{certGrayLight}{HTML}{F2F4F4}

\definecolor{grondblue}{RGB}{41, 128, 185}
\definecolor{osdorange}{RGB}{211, 84, 0}
\definecolor{nodebg}{RGB}{248, 249, 250}

\newif\ifsubmission
\submissionfalse
\newcommand{\appref}[1]{%
  \ifsubmission
  the extended version~\cite{banfiGithubBridge}%
  \else
  Sec.~\ref{#1} of the Appendix%
\fi}

\newcommand{\Hdec}{H_{\text{dec}}}
\newcommand{\Hlog}{H_{\text{log}}}
\newcommand{\Haug}{H_{\text{aug}}}
\newcommand{\cost}{\text{cost}}
\newcommand{\LB}{\mathrm{LB}}
\newcommand{\UB}{\mathrm{UB}}

\begin{document}
\title{Certified Non-Degenerate Maximum-Likelihood \\ Benchmarking of Quantum LDPC Decoders}

\author{
  \IEEEauthorblockN{Michele Banfi$^{*\dagger}$, Marco Ferrari$^{\dagger}$, Antonino Favano$^{*}$, Alberto Tarable$^{\dagger}$, and Luca Barletta$^{*\dagger}$}
  $^{\dagger}$Consiglio Nazionale delle
  Ricerche, Italy. E-mail: $\{$alberto.tarable, marcopietro.ferrari$\}$@cnr.it\\
  $^{*}$Politecnico di Milano, Milan, 20133, Italy. E-mail:  $\{$michele.banfi, antonino.favano, luca.barletta$\}$@polimi.it
}

\maketitle

\begin{abstract}
  The measured logical error rate of a quantum low-density parity-check (QLDPC) code depends on the decoding algorithm.
  We develop a framework that certifies non-degenerate maximum-likelihood (ML) decoding outcomes on individual sampled shots under circuit-level noise, providing a common reference for comparing decoders and codes.
  Upper and lower bounds on the minimum correction cost in the true and competing logical classes establish success, failure, or an undecided outcome.
  Branch-and-bound search refines these intervals while retaining valid bounds under resource limits.
  Counting the outcomes bounds the empirical ML logical error rate, with interval width equal to the undecided fraction.
  Results for two bivariate bicycle codes show that best-first ordered statistics decoding (BF-OSD) tracks the non-degenerate ML benchmark closely while leaving some ML-correctable shots.
\end{abstract}

\section{Introduction}

The error rate performance of a QLDPC code under practical decoding schemes is typically assessed by Monte Carlo simulation.
On the other hand, Monte Carlo simulation typically cannot establish the performance of the same code under a specified maximum-likelihood (ML) rule, whose evaluation is
generally
computationally demanding \cite{iyer2015hardness}.
Belief propagation (BP) with ordered statistics decoding (OSD) is the current state-of-the-art for QLDPC decoding \cite{panteleev2021degenerate,roffe2020decoding}, and refinements continue to improve measured logical error rates and efficiency \cite{banfi2026bfosd}.
However, a search that finds no better correction does not establish optimality unless it also bounds the unexplored candidates.

A \emph{certificate} is a set of verified bounds that proves a decoding outcome for a sampled shot, even if some candidate corrections remain unexplored.
The term \emph{cost} refers to the likelihood of an error: a more likely error has a lower cost.
We use such certificates to establish the outcome of \emph{non-degenerate} ML decoding: selecting the most probable error pattern under the specified detector error model and returning its logical effect, that is, the effect onto the logical qubits, referred to here as \emph{class}. Two different physical errors can perform the same logical operation, i.e. they are part of the same class.
Non-degenerate ML differs from degenerate ML, which maximizes the total probability of a class.
For each sampled shot, bounds establish whether the most probable correction belongs to the true class, or leave the outcome undecided.
Shot aggregation yields lower and upper bounds on the empirical non-degenerate ML error rate.
The bounds provide a reference for comparing decoders and identify decoder failures on which this ML rule would succeed.

The framework applies to any linear code specified by the detector and logical parity-check matrices and to any decoder that supplies candidate corrections.
The main text develops the certification principle and the search used to establish its bounds.
Computational accelerations are described in \appref{app:accelerations}.
We apply the benchmark to bivariate bicycle (BB) codes under circuit-level noise \cite{bravyi2024high}.

\section{The Certification Procedure}
\label{sec:overview}

For each simulated shot, i.e. a sampled event and its correction, the sampled error identifies the true logical class.
Certification uses two cost intervals, containing the minimum cost in the true class and the minimum over competing classes, respectively. Both intervals can be narrowed by lowering the upper bound or increasing the lower bound.
A verified lower-cost correction improves an upper bound whereas demonstrating that lower-cost candidates do not exist improves a lower bound.
The shot is certified when the two cost intervals are separated by the required margin.

The following list summarizes the procedure:

\begin{enumerate}[leftmargin=*]
  \item \textbf{Sample a shot.} Record the syndrome and true logical class in each sector $X$ and $Z$ (Sec. \ref{sec:setup}).
  \item \textbf{Initialize cost intervals.} Use verified corrections and valid lower bounds (Sec. \ref{sec:certs}).
  \item \textbf{Refine unresolved intervals.} Search by branch and bound, retaining valid bounds at resource limits (Sec. \ref{sec:engine}).
  \item \textbf{Compare the intervals.} Record certified success, certified failure, or an undecided outcome (Sec. \ref{sec:decision}).
  \item \textbf{Count the outcomes.} Combine $X$ and $Z$ decoding outcomes and bound the empirical ML error rate (Sec. \ref{sec:aggregation}).
\end{enumerate}

The comparison in Step~4 follows every bound improvement, so a shot can be classified without branch-and-bound when its initial bounds already separate the classes.
The complete workflow, including computational accelerations, appears in \appref{app:accelerations}.
Section~\ref{sec:example} illustrates this procedure through an example.

\section{Noise Model and Decoding Objective}
\label{sec:setup}

\subsection{Faults, Syndromes, and Costs}

A shot is one simulated execution of the memory circuit, including $N_c$ rounds of syndrome extraction.
We use the full circuit-level noise model of \cite{bravyi2024high,banfi2026bfosd}, which includes faults in gates, idling, preparation, and measurement.
A fault can propagate through subsequent gates, and its observable effect is recorded by detectors that compare measurement outcomes.
The $X$ and $Z$ sectors are decoded separately; a shot fails if either sector fails. The following notation describes one sector.

As in \cite{banfi2026bfosd}, decoding uses a matrix $\Hdec\in\mathbb F_2^{m\times n}$ whose $m$ rows index detectors and whose $n$ columns index fault mechanisms.

Let $e\in\mathbb F_2^n$ be the sampled error pattern, with $e_j=1$ when
mechanism $j$ occurs. The observed syndrome is $s=\Hdec e$.
The decoder seeks a candidate $x\in\mathbb F_2^n$ that explains the
observations, so $x$ must satisfy
\begin{equation}
  \Hdec x=s.
  \label{eq:coset}
\end{equation}
Equation~\eqref{eq:coset} may have many solutions, ranked by the decoder according to their probability under the noise model.

For certification, we use the column-level model in which mechanisms occur independently with probabilities $0<q_j<1/2$.
The prior log-likelihood ratio of mechanism $j$ is
\begin{equation}
  w_j=\log\frac{1-q_j}{q_j}>0,
  \label{eq:weights}
\end{equation}
and it measures the cost of including that mechanism in a candidate.
The total candidate cost is the sum over its selected mechanisms:
\begin{equation}
  \cost(x)=\sum_{j=1}^{n}w_jx_j.
  \label{eq:cost}
\end{equation}
Independence gives
\begin{equation}
  -\log P(x)=-\sum_j\log(1-q_j)+\cost(x)
\end{equation}
where $P(x)$ is the probability of the fault pattern $x$.
The first term in the right-hand side (RHS) is the same for every candidate, so minimizing \eqref{eq:cost} selects one of the most probable patterns satisfying \eqref{eq:coset}. We use natural logarithms and express costs in nats.
These costs come from the specified noise model and remain fixed during certification.

\subsection{Logical Classes and Recovery Conditions}

The logical matrix $\Hlog\in\mathbb F_2^{k\times n}$, where $k$ is the number of logical operators, records the logical effect of each mechanism. For an observed syndrome $s$, define
\begin{align*}
  \mathcal X(s)
  &=\{x\in\mathbb F_2^n:\Hdec x=s\},\\
  \mathcal X_\lambda(s)
  &=\{x\in\mathcal X(s):\Hlog x=\lambda\},
  \qquad \lambda\in\mathbb F_2^k.
\end{align*}
Each $\mathcal X_\lambda(s)$ contains the syndrome-consistent corrections with logical class $\lambda$. These sets partition $\mathcal X(s)$, although some may be empty. The sampled error defines the true class $\lambda_0=\Hlog e$.
A correction succeeds when $x\in\mathcal X_{\lambda_0}(s)$, equivalently when $\Hlog(x\oplus e)=0$.
Thus, the syndrome-consistent patterns fall into at most $2^k$ nonempty logical classes. Distinct physical errors can have the same action on the encoded state when they differ by a stabilizer.
This property is called \emph{degeneracy} \cite{poulin2008iterative}.
In the detector model, patterns with the same syndrome and logical class therefore have the same recovery outcome, even when their mechanisms differ.

\subsection{The Non-Degenerate ML Reference}

For each logical class, define the cost of its cheapest representative:
\begin{equation}
  m_\lambda(s)
  =\min_{x\in\mathcal X_\lambda(s)}\cost(x),
  \label{eq:classmin}
\end{equation}
where the minimum of an empty class is $+\infty$.
The true class is nonempty because it contains the sampled error $e$.

The non-degenerate ML decoder returns the class of a globally
minimum-cost pattern:
\begin{equation}
  \hat\lambda\in
  \arg\min_{\lambda\in\mathbb F_2^k}m_\lambda(s).
  \label{eq:nondegml}
\end{equation}
An exhaustive search of the syndrome-consistent candidates under the prior cost metric implements this rule.
Degenerate ML instead selects the class with the greatest \emph{total} probability, summed over its representatives.
The two outputs can differ; minimum representative cost does not in general minimize logical error probability, i.e., the total class probability.

For the rest of the analysis and to ease the notation, the syndrome $s$ is fixed and is omitted from $m_\lambda(s)$.
The comparison uses the true-class minimum $m_{\lambda_0}$ and the minimum over competing classes, $m_w=\min_{\lambda\ne\lambda_0}m_\lambda$.
When these minima differ, non-degenerate ML fails iff $m_w<m_{\lambda_0}$.
If the minima are equal, the outcome depends on tie-breaking.
We leave such shots undecided, so the error-rate bounds apply to any fixed tie-breaking rule.

\section{Certification from Cost Intervals}
\label{sec:certs}

Exact minima can be expensive to compute. Instead, we maintain one interval containing the true-class minimum and another containing the cheapest competing-class minimum:
\begin{equation}
  \ell_0\le m_{\lambda_0}\le u_0,\qquad
  \ell_w\le m_w\le u_w.
  \label{eq:cost-intervals}
\end{equation}
The subscript identifies the true class ($0$) or a competitor ($w$).

\subsection{Upper Bounds from Feasible Corrections}
\label{sec:upper-bounds}

A feasible correction is a \emph{witness}: its cost bounds from above the minimum over any set containing it.
The sampled error $e$ supplies $u_0=\cost(e)$; a cheaper true-class witness improves this bound.
A wrong-class witness supplies $u_w$, with $u_w=+\infty$ if none is available.
Witness construction is developed in \appref{app:accelerations}.

\subsection{Lower Bounds from Parity Constraints}
\label{sec:lower-bounds}

To bound a minimum from below, we identify costs that \emph{every} feasible correction must incur. Since all weights are positive, zero is always a valid lower bound. The observed detector syndrome often gives a better one.

Let $D_r=\{j:(\Hdec)_{rj}=1\}$ be the set of mechanisms that flip detector $r$. When $s_r=1$, a correction must select an odd number of mechanisms from $D_r$. In particular, it must select at least one and pay at least $\min_{j\in D_r}w_j$. This argument applies regardless of the logical class of the correction.

One mechanism may flip two detectors, so their individual minimum costs cannot generally be added: the same mechanism might satisfy both at once.
We obtain a \emph{detector bound} by choosing a subset $R$ of triggered detector rows ($s_r=1$) whose mechanism sets $D_r$ are pairwise disjoint.
Every syndrome-consistent correction must then select at least one distinct mechanism from each set, giving for every $x \in \mathcal{X}(s)$
\begin{equation}
  \cost(x)\ge h_R:=\sum_{r\in R}\min_{j\in D_r}w_j.
  \label{eq:detector-bound}
\end{equation}
For example, two triggered detectors with disjoint mechanism sets whose cheapest weights are $1$ and $2$ require total cost at least $3$. If their sets overlap, adding $1+2$ may be invalid. Any pairwise-disjoint choice of $R$ gives a sound bound; finding the strongest such choice is unnecessary for certification.

Because this bound uses the syndrome alone, it can initialize both $\ell_0$ and $\ell_w$ in \eqref{eq:cost-intervals}. Logical-class constraints may raise either bound further.
The search below applies the same reasoning after some mechanisms have been fixed: it charges for selected mechanisms and bounds the cost still required by the remaining parity equations.

\subsection{Certification Conditions}
\label{sec:decision}

\begin{proposition}[Verdict soundness]
  \label{prop:verdict}
  For a fixed margin $\varepsilon>0$, $\ell_w>u_0+\varepsilon$ certifies success and $u_w<\ell_0-\varepsilon$ certifies failure.
  Otherwise, the sector remains undecided.
\end{proposition}
\begin{proof}
  Success follows from $m_w\ge\ell_w>u_0\ge m_{\lambda_0}$; failure follows from $m_w\le u_w<\ell_0\le m_{\lambda_0}$.
\end{proof}

For success, we exhibit a true-class correction costing at most $u_0$ and prove that every competitor costs at least $\ell_w>u_0+ \varepsilon$.
For failure, we exhibit a competitor costing at most $u_w$ and prove that every true-class correction costs at least $\ell_0>u_w+\varepsilon$. These comparisons need bounds on the minima, not the exact minima themselves. If the intervals are not separated by the required margin, we keep the result undecided and may refine the bounds further.

\section{Branch-and-Bound Refinement}
\label{sec:engine}

The detector bound~\eqref{eq:detector-bound} alone may leave the two intervals overlapping.
An interval can be narrowed by improving the lower bound. The lower bound search can be made more efficient by branch-and-bound: a logical class can be divided into smaller sets of corrections. If the lower bound of a set is at least the current upper bound, searching that set cannot yield a cheaper correction, so the set can be discarded.

The baseline runs this search for each logical class in \eqref{eq:classmin}. Write $L_\lambda$ for the lower bound of class $\mathcal{X}_\lambda(s)$ and $U_\lambda$ for the cost of its cheapest candidate verified so far.
If no candidate has been found, $U_\lambda=+\infty$. At every stage $L_\lambda\le m_\lambda\le U_\lambda$.
The two intervals compared in Proposition~\ref{prop:verdict} are
\begin{equation}
  \begin{aligned}
    \ell_0&=L_{\lambda_0}, & u_0&=U_{\lambda_0},\\
    \ell_w&=\min_{\lambda\ne\lambda_0}L_\lambda,
    & u_w&=\min_{\lambda\ne\lambda_0}U_\lambda.
  \end{aligned}
  \label{eq:class-bounds}
\end{equation}
In particular, proving success requires a lower bound for \emph{every} competing class. The smallest of those bounds is the guarantee for the union of all competitors.

\subsection{Search Nodes and Their Bounds}

Each class can be represented as  a separate search tree. At its root, no mechanism has been assigned.
A node $v$ fixes some mechanisms to zero (absent) or one (present) and represents the subset of class-$\lambda$ corrections consistent with such choices.
Its children fix more mechanisms, dividing that subset into smaller pieces.

The mechanisms fixed to one have already incurred  a cost, say, $g(v)$.
Their contributions to the target detector syndrome and logical class are removed by XOR. The remaining, or \emph{residual}, parity equations specify what the unassigned mechanisms must still do.
For example, if the residual syndrome value of a detector is 1 (i.e., the detector is still unsatisfied), an odd number of the unassigned mechanisms affecting that detector must be selected.
Applying the disjoint-support bound \eqref{eq:detector-bound} to these residual detector rows gives a lower bound $h(v)$ on the additional cost.
Thus,
\begin{equation}
  B(v)=g(v)+h(v)
  \label{eq:node-bound}
\end{equation}
is no greater than the cost of any feasible completion of $v$.
If an unsatisfied detector or logical row has no unassigned mechanism, the node is infeasible and can be discarded.

\subsection{Branching and Pruning}

The \emph{frontier} contains nodes awaiting exploration.
We expand a node with the smallest $B(v)$ first, as in best-first or A$^\star$ search \cite{hart1968astar}. At an unfinished node, choose a parity row (detector or logical) whose residual right-hand side, after substituting the assigned variables, is 1, with unassigned columns $j_1,\ldots,j_t$. Every feasible completion must select at least one of them. Make child $a$ by fixing $x_{j_a}=1$ and $x_{j_1}=\cdots=x_{j_{a-1}}=0$; later columns remain unassigned.
Each feasible completion goes to exactly one child, determined by its first selected column. No feasible correction is lost or counted in two children.

There are two ways to finish exploration on a node:
\begin{itemize}
  \item If every residual parity is zero, setting all unassigned mechanisms to zero gives the cheapest completion of this node because all weights are positive. We verify the detector and logical parities of the completed correction before using its cost to reduce $U_\lambda$.
  \item If $B(v)\ge U_\lambda$, no completion can cost less than the verified candidate. We \emph{prune} the node without enumerating its children.
\end{itemize}
The search can stop as soon as the intervals certify a verdict; it need not finish every class search.

\subsection{Resource Limits and Unresolved Outcomes}
\label{sec:early-stop}

A time or memory budget may stop a search before all nodes are finished. Bounds remain valid if we account for every unfinished part of the class. Let $F_\lambda$ be the smallest $B(v)$ among nodes still on the frontier. If a cap removes unfinished nodes from that frontier, retain their bounds and let $D_\lambda$ be the smallest of those retained values. An empty collection has value
$+\infty$. Then
\begin{equation}
  L_\lambda=\min\{U_\lambda,F_\lambda,D_\lambda\}
  \label{eq:frontier-bound}
\end{equation}
is a valid class lower bound. To see why, every class-$\lambda$ correction lies in one of three places: a completed or safely pruned region accounted for by the verified incumbent $U_\lambda$, an active node bounded by $F_\lambda$, or an unfinished removed node bounded by $D_\lambda$. The smallest of these bounds therefore applies to the class minimum. A valid lower bound already proved for the class can be kept by taking its maximum with this expression.

Algorithm~\ref{alg:branch-bound} gives the class search in procedural form \cite{banfi2026qldpc}.

\begin{algorithm}[!t]
  \centering
  \footnotesize
  \begin{algorithmic}
    \STATE \textbf{Input:} $s,\lambda,w$, budget, verified $U_\lambda$
    \STATE $Q\gets\{\text{root for }(s,\lambda)\}$; $D_\lambda\gets+\infty$
    \WHILE{$Q\ne\varnothing$ and budget remains}
    \STATE Pop $v$ of minimum $B(v)$ from $Q$
    \STATE \textbf{if} $v$ is infeasible or $B(v)\ge U_\lambda$, \textbf{continue}
    \IF{all residual parities are zero}
    \STATE Complete with zeros to obtain $x$; verify its parities
    \STATE \textbf{if} valid, $U_\lambda\gets\min\{U_\lambda,\cost(x)\}$
    \ELSE
    \STATE Choose a residual-one row with columns $j_1,\ldots,j_t$
    \FOR{$a=1,\ldots,t$}
    \STATE Fix $x_{j_a}=1$, earlier $x_{j_b}=0$; bound child $v_a$
    \STATE \textbf{if} feasible and $B(v_a)<U_\lambda$, keep $v_a$ in
    $Q$, or retain $B(v_a)$ in $D_\lambda$ if $Q$ is full
    \ENDFOR
    \ENDIF
    \ENDWHILE
    \STATE $L_\lambda\gets\min\{U_\lambda,\min_{v\in Q}B(v),D_\lambda\}$
    \STATE \textbf{Return:} $[L_\lambda,U_\lambda]$
  \end{algorithmic}
  \caption{Bounded branch-and-bound search for one logical class.}
  \label{alg:branch-bound}
\end{algorithm}

The retained bound of a removed node still matters even if the active frontier becomes empty. Discarding that bound would wrongly treat an unexplored region as resolved. If a resource limit leaves the comparison open, we record \textsc{budget}. If the two minima are resolved but differ by at most $\varepsilon$, we record \textsc{tie}.
Both outcomes remain undecided for the empirical error-rate bounds. The guarantees assume valid numerical bounds; solver arithmetic is discussed in \appref{app:relax}.

\section{Empirical Error-Rate Bounds}
\label{sec:aggregation}

Once each sector has a verdict, a shot is a certified failure if either sector fails, a certified success if both sectors succeed, and undecided otherwise. Counting those outcomes gives the empirical benchmark.

\begin{proposition}
  \label{prop:bracket}
  Over $N$ sampled shots, the empirical non-degenerate ML logical
  error rate $\widehat P_L^{\mathrm{ML}}$ satisfies
  \begin{equation}
    \begin{aligned}
      \LB=\frac{\#\{\text{cert. fail}\}}{N}
      \le\widehat P_L^{\mathrm{ML}}
      \le1-\frac{\#\{\text{cert. success}\}}{N}=\UB.
    \end{aligned}
    \label{eq:bracket}
  \end{equation}
\end{proposition}

The lower bound counts only proved failures. The upper bound treats every undecided shot as a possible failure. Consequently, the width is exactly the undecided fraction:
\[
  \UB-\LB=\frac{N_{\mathrm{undecided}}}{N}.
\]
The quantities $\LB$ and $\UB$ bound an empirical \emph{error rate}; the lowercase bounds $\ell_0,u_0,\ell_w,u_w$ concern per-sector \emph{costs}. More computation may narrow the rate interval by deciding more shots, without changing the sampled data.

The interval \eqref{eq:bracket} accounts only for unfinished computation on the sampled shots. To bound the population probability $P_L^{\mathrm{ML}}$ that a random shot is an ML failure, let the $N$ shots be independent, with a number $N$ fixed in advance, and let $F$ and $S$ be the numbers of certified failures and certified successes.
Every certified failure is a true ML failure, so $F$ is at most a $\operatorname{Binom}(N,P_L^{\mathrm{ML}})$ count; likewise $S$ is at most a $\operatorname{Binom}(N,1-P_L^{\mathrm{ML}})$ count.
Because the Clopper--Pearson lower limit~\cite{clopper1934use} $\mathrm{CP}(x,N;\alpha) =\mathrm{Beta}^{-1}(\alpha;\,x,\,N-x+1)$ increases with $x$, applying it to these certified counts gives valid, conservative bounds:
\begin{equation}
  \mathrm{CP}(F,N;\tfrac{\alpha}{2})
  \;\le\; P_L^{\mathrm{ML}} \;\le\;
  1-\mathrm{CP}(S,N;\tfrac{\alpha}{2}),
  \label{eq:population}
\end{equation}
each side holding with probability at least $1-\alpha/2$, hence jointly with probability at least $1-\alpha$ by a union bound.

The upper endpoint in \eqref{eq:population} combines the uncertainties of undecided shots and sampling.
It may therefore exceed the measured error rate of another decoder without establishing its superiority.
Further certification may reduce the undecided fraction, whereas sampling uncertainty remains even when all outcomes are certified. One could take the performance of a practical decoder as an upper bound.
For example, BF-OSD \cite{banfi2026bfosd} could be used as an upper bound for the $[[72, 12, 6]]$ code.

\section{Worked Example}
\label{sec:example}

A small algebraic detector model illustrates how the preceding procedure establishes success, failure, and empirical error-rate bounds.

\paragraph*{Sampled error and cost}
The example consists of two detectors and five mechanisms:
\begin{equation}
  \Hdec=
  \begin{bmatrix}1&1&0&0&0\\0&0&1&1&0
  \end{bmatrix},
  \qquad w=(1,4,2,5,2).
  \label{eq:toy-detectors}
\end{equation}
Set $q_j=(1+\exp(w_j))^{-1}$, so $w$ contains exactly the prior log-odds of the independent mechanisms.

Consider the sampled error $e^{(\mathrm S)}=(0,1,1,0,1).$
Mechanisms $2$, $3$, and $5$ fire. Their total cost is $4+2+2=8$, and the observed syndrome is $s=(1,1)$. Any correction must satisfy $x_1\oplus x_2=1$ and $x_3\oplus x_4=1$. These constraints admit eight patterns.

\paragraph*{Logical-class assignment}
Choose two logical bits with matrix
\begin{equation}
  \Hlog=
  \begin{bmatrix}0&1&0&0&1\\0&0&0&1&0
  \end{bmatrix}.
  \label{eq:toy-logicals}
\end{equation}
The class of a candidate is $\lambda = (x_2\oplus x_5,x_4)$.
The sampled error has class $\lambda_0 =(1\oplus1,0)=00$.
Column~5 changes the first logical bit despite having no detector signature, so it must be retained.

\paragraph*{True-class upper bound}
The sampled error gives $u_0=8$. The correction $x^{(0)}=(1,0,1,0,0)$ has the same syndrome and class $00$, but costs $1+2=3$, reducing $u_0$ to $3$. No wrong-class witness is retained, so $u_w=+\infty$. A competing witness of cost $5$ would still cost more than $x^{(0)}$, despite being cheaper than the sampled error.

\paragraph*{Detector lower bound}
Detector~1 requires a mechanism from $\{1,2\}$, costing at least $1$; detector~2 requires one from $\{3,4\}$, costing at least $2$.
These disjoint sets imply cost at least $3$ for every correction, hence
$
[\ell_0,u_0]=[3,3], $ and $ [\ell_w,u_w]=[3,+\infty].
$
The true-class minimum is now known, but the competing-class minimum is not yet separated from it.

\paragraph*{Success certificate}
For classes $01$ and $11$, $x_4=1$ costs $5$ and detector~1 still requires cost at least $1$, so both class minima are at least $6$.
Only class $10$ requires refinement. Its constraints include $x_2\oplus x_5=1$ and $x_4=0$. In order to match both the syndrome $s$ and class $\lambda = 10$ it expands its root on detector~1:
\begin{enumerate}
  \item $x_1=1$ forces $x_2=0$ and $x_5=1$. Together with
    detector~2, the bound is $B=1+2+2=5$.
  \item $x_1=0$ forces $x_2=1$ and $x_5=0$. Detector~2 adds at
    least $2$, giving $B=4+2=6$.
\end{enumerate}
Thus $L_{10}=5$ and $\ell_w=\min\{5,6,6\}=5$. For $\varepsilon=10^{-9}$, $5>3+\varepsilon$ certifies success without a wrong-class witness.

\paragraph*{Failure certificate}
With the same model, sample $e^{(\mathrm F)}=(1,0,1,0,1)$, of cost $5$ and true class $10$. The correction $x^{(0)}$ is now a wrong-class witness of cost $3$. The detector bounds give intervals $[3,5]$ and $[3,3]$. Expanding the true-class root yields the same bounds $5$ and $6$ above, raising $\ell_0$ to $5$. Since $3<5-\varepsilon$, the sector is a certified failure.

\paragraph*{Budget limits and ties}
Stopping shot S before expansion leaves $[3,3]$ and $[3,+\infty]$, so it remains undecided. After expansion, even if a cap $C=3.25$ removes both class-$10$ branches, retaining their bounds $5$ and $6$ preserves the success certificate.
An exact tie is different: changing $w_2$ to $1$ gives classes $00$ and $10$ the same minimum cost $3$.
No amount of refinement separates them; the outcome remains undecided.

\paragraph*{Empirical bounds for three shots}
Return to the original weights and consider the sampled errors $(e^{(\mathrm S)},e^{(\mathrm F)},e^{(\mathrm S)})$. In the other sector, take the same toy model with zero sampled error on each shot.
There, the zero correction has true-class cost $0$, while every wrong-class correction needs at least one mechanism and costs at least $1$.
That sector is therefore certified successful on all three shots.
Now, run the root expansions described above on the first two shots, but stop at the detector bounds on the third. The outcomes are success, failure, and undecided. Hence
$
\LB=\frac13, \UB=\frac23.
$
Refining the third shot certifies success and collapses the interval
to $[1/3,1/3]$.

\setlength{\textfloatsep}{6pt plus 2pt minus 2pt}
\begin{figure}[!t]
  \centering
  \includegraphics[width=\linewidth]{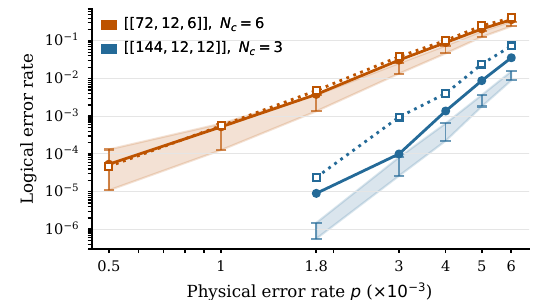}
  \caption{Non-degenerate ML logical error rates for $[[72,12,6]]$,
    $N_c=6$ (orange), and $[[144,12,12]]$, $N_c=3$ (blue). Shaded bands
    show joint 95\% confidence intervals \eqref{eq:population}; solid
  lines show BF-OSD and dotted lines show BP\,+\,OSD-CS.}
  \label{fig:results}
\end{figure}

\section{Results}
\label{sec:results}

We apply the same benchmark to two BB codes under the full circuit-level noise model of \cite{bravyi2024high}: $[[72,12,6]]$ with $N_c=6$ syndrome-extraction rounds and $[[144,12,12]]$ with $N_c=3$.
Rates refer to complete shots.
In both experiments, BF-OSD \cite{banfi2026bfosd} is evaluated against the certified benchmark; its corrections are not used in the certification search. BP\,+\,OSD-CS \cite{roffe2020decoding} and the beam-search decoder Tesseract \cite{aghababaie2025tesseract} may supply candidate corrections that, after verification, initialize the upper cost bounds.

Figure~\ref{fig:results} shows that the BF-OSD error rate stays close to the non-degenerate ML benchmark on both codes, but lies above it at most sampled rates.
The upper ML limit can exceed the decoder point estimates because unresolved near-ties count as possible failures and sampling uncertainty widens the interval; non-degenerate ML itself is not guaranteed to minimize logical error.
From $p=3\times10^{-3}$ upward, the gap exceeds the statistical uncertainty of both rates. Thus BF-OSD misses corrections that this ML rule would make, although its error rate remains of the same order as the benchmark.
At the lowest rates, the ML interval is widened mainly by near-ties: shots whose true-class and competing minima differ by at most $\varepsilon$ remain undecided, no matter how much computation is spent.

Only at $p=3\times10^{-3}$ BF-OSD sample contain too few failures to resolve the gap. Together, the experiments indicate that BF-OSD operates near the non-degenerate ML reference on the smaller code, while the certificates identify a margin for improvement on the larger one.

\vspace{-0.5pt}

\section{Conclusion}
\label{sec:conclusion}

We have presented a procedure to certify non-degenerate maximum likelihood decoding for any linear quantum low density parity check code.
The framework certifies whether this decoding rule succeeds or fails on individual simulated shots and bounds its empirical logical error rate when some outcomes remain undecided.
Aggregating those certificates gives an empirical reference for comparing practical decoders under a fixed noise model. On both bivariate bicycle codes, best-first ordered statistics decoding performs close to that reference, although the certificates identify room for improvement on some shots.
Stronger bounds at difficult noise rates would make this comparison more precise.

\section*{AI Statement}
Generative AI tools were used to polish the writing.

\setlength{\textfloatsep}{\savedtextfloatsep}
\bibliographystyle{IEEEtran}
\bibliography{bibliography}

\ifsubmission\else
\clearpage


\appendix[Computational Methods]
\input{fig/styles}

This appendix develops the computational refinements used to make the certification procedure practical. The aim is to tighten cost intervals with less work while preserving the verdict conditions of Proposition~\ref{prop:verdict} and the empirical bracket in \eqref{eq:bracket}. We first describe witness initialization and search reductions, then establish the conditions for cluster decomposition and explain the optimization models. Each reduction must retain every correction relevant to a verdict or bound the cost of those it excludes; the final subsection discusses the numerical qualifications for solver-based bounds.

\subsection{Computational Accelerations}
\label{app:accelerations}

The main text describes a separate search for each logical class. This section explains how we obtain valid cost bounds with less work. We first improve the verified corrections used as upper bounds, then replace the many competing-class searches by $k$ simpler problems. Cluster decomposition and optimization bounds provide further ways to resolve difficult sectors. Figure~\ref{fig:roadmap-full} places all the steps in the full workflow.
\input{fig/roadmap_full}

\subsubsection{Witness Initialization}
\label{sec:anchor}

A low-cost verified correction provides a tight upper bound from the start of the search. For the true class, the sampled error $e$ always supplies an initial witness. In our experiments, we also construct an \emph{anchor} correction by running a short BP pass followed by zero-order ordered statistics decoding (OSD-0) \cite{roffe2020decoding} on the augmented parity-check matrix $\Haug=[\Hdec;\Hlog]$ with target $(s,\lambda_0)$. Briefly, OSD-0 uses the BP reliability ranking to select a full-rank set of columns, solves for the corresponding correction bits, and sets all remaining bits to zero without searching higher-order patterns. We initialize $u_0$ as the smaller of $\cost(e)$ and the verified anchor cost.

Candidate corrections from other decoders can similarly serve as witnesses: after verifying their detector and logical parities, their costs can lower $u_0$ (if in class $\lambda_0$) or $u_w$ (if in a competing class). Because a tighter initial witness allows more shots to be certified within a fixed computational budget, the decoder being benchmarked (here, BF-OSD) is never used to seed the certification search. Instead, only auxiliary decoders (such as BP\,+\,OSD-CS and Tesseract in Sec.~\ref{sec:results}) supply candidate witnesses.

\subsubsection{Reduction to $k+1$ Constrained Problems}
\label{sec:reduction}

The baseline procedure in Sec.~\ref{sec:engine} searches each of the up to $2^k$ logical classes separately. To bound the competing-class minimum $m_w$, however, this is unnecessary: every competing correction differs from the true class $\lambda_0$ in at least one of the $k$ logical bits. We can therefore replace the $2^k-1$ competing-class searches with $k$ searches, one per logical bit. For each bit $i \in \{1,\ldots,k\}$, let $\mu_i(s)$ be the minimum cost of a syndrome-consistent correction whose $i$-th logical bit is flipped relative to $\lambda_{0,i}$:
\begin{equation}
  \mu_i(s)=\min\{\cost(x): x\in\mathbb{F}_2^n,\ \Hdec x=s,\ (\Hlog x)_i=\lambda_{0,i}\oplus1\}.
  \label{eq:wrongbit}
\end{equation}
Because the remaining $k-1$ logical bits are left unconstrained, the feasible set of~\eqref{eq:wrongbit} is the union of all competing classes that disagree with $\lambda_0$ on bit $i$.

The cheapest competing-class cost $m_w$ is therefore the minimum over the $k$ wrong-bit minima $\mu_i(s)$.

\begin{lemma}[Union reduction]
  \label{lem:union}
  $\displaystyle m_w=\min_{i=1,\ldots,k}\mu_i(s)$.
\end{lemma}

\begin{proof}
  Every wrong-class candidate differs from the true class in some coordinate $i$, so it is feasible for at least one wrong-bit problem. Conversely, every candidate feasible for a wrong-bit problem belongs to a wrong class. The union of these feasible sets is therefore exactly the set of competing-class candidates, and minimizing over the union gives the stated identity.
\end{proof}

Certification now needs $k+1$ constrained searches. The true-class search fixes all logical bits to $\lambda_0$; wrong-bit search $i$ fixes only bit $i$ to its opposite value. They use the same detector rows but different active rows of $\Hlog$. For example, with $k=12$, this is $13$ searches per sector rather than as many as $4096$ class searches.

Let $\ell_i$ and $u_i$ be lower and upper bounds on the wrong-bit minimum $\mu_i$, so $\ell_i\le\mu_i\le u_i$. The bounds needed by the certificate are
\begin{equation}
  \ell_w=\min_i\ell_i,\qquad u_w=\min_i u_i.
  \label{eq:wrong-bound-combination}
\end{equation}
The minimum across the wrong-bit lower bounds applies to all competitors, even when the wrong-bit feasible sets overlap.
Therefore, success requires a sufficiently high lower bound for every bit; failure needs only one sufficiently cheap competing witness.

\subsubsection{A Single Syndrome-Only Search}
\label{sec:unified}

The $k+1$ searches in Sec.~\ref{sec:reduction} still solve the same detector equations $\Hdec x = s$ repeatedly. We can instead run a single best-first search that enforces only $\Hdec x = s$ and leaves all logical rows unconstrained, reading off the class $\Hlog x$ of each completed correction.

Because best-first search expands a frontier node minimizing $B(v)=g(v)+h(v)$, and any node with zero residual syndrome is completed by setting all remaining unassigned mechanisms to zero (incurring zero additional cost), completed corrections are enumerated in nondecreasing order of cost. Consequently, the first completed correction $x^*$ attains the global minimum cost across all classes, $m_* = \cost(x^*) = \min_\lambda m_\lambda$, and identifies a globally cheapest class $\lambda_* = \Hlog x^*$.

To certify that $\lambda_*$ is strictly cheaper than every competing class by at least $\varepsilon$, we continue expanding nodes until every remaining frontier node (and any node removed under a memory cap, as in~\eqref{eq:frontier-bound}) satisfies $B(v) > m_* + \varepsilon$. Note that when a node reaches zero residual syndrome and yields a correction $x$, Algorithm~\ref{alg:branch-bound} terminates that branch rather than assigning further mechanisms. Any omitted completion along that branch has the form $x \oplus \Delta x$ with $\Delta x \neq 0$ and $\Hdec \Delta x = 0$; since all weights are positive, adding $\Delta x$ increases the cost by at least $w_{\min} := \min_j w_j$, so any such extension costs at least $m_* + w_{\min}$.

The stopping condition must also exclude near-optimal corrections in other logical classes. The following lemma establishes that separation.

\begin{lemma}[Single-search verdict]
  \label{lem:unified}
  Suppose $w_{\min}>\varepsilon$ and the search stops as above with no completed correction outside $\lambda_*$ costing at most $m_*+\varepsilon$. Then every class other than $\lambda_*$ has minimum greater than $m_*+\varepsilon$.
\end{lemma}

\begin{proof}
  Consider a correction outside $\lambda_*$. There are three cases. If it was completed, its cost exceeds $m_*+\varepsilon$ by assumption. If it lies below a frontier or retained node, the bound at that node exceeds $m_*+\varepsilon$. Otherwise, it extends a completed correction and costs at least $m_*+w_{\min}>m_*+\varepsilon$.
\end{proof}

The class of the first minimum-cost correction determines the verdict once the lemma separates it from every other class.
If $\lambda_*=\lambda_0$, the lemma gives $u_0=m_*$ and $\ell_w>m_*+\varepsilon$, which certifies success by Proposition~\ref{prop:verdict}. If $\lambda_*\ne\lambda_0$, it gives $u_w=m_*$ and $\ell_0>m_*+\varepsilon$, which certifies failure.

A completed correction of another class within $\varepsilon$ of $m_*$ leaves the sector undecided, as does a resource limit reached before the frontier clears $m_*+\varepsilon$.

\subsubsection{Additional Bound Improvements}
\label{sec:accelerations}

The remaining methods aim to tighten the bounds or reduce search work. Cluster decomposition divides fired detectors into groups once we have proved that no correction below a chosen cost can connect them. We then solve smaller syndrome problems and combine their witnesses, under the conditions proved in \appref{app:clusters}.

For sectors that remain unresolved, linear and integer programming relaxations solved with HiGHS~\cite{huangfu2018highs} can tighten the lower bounds or identify cheaper witnesses. Furthermore, integer feasibility tests can certify a decoding sector directly without computing exact class minima, while parity cuts and cost-based column filtering reduce the size of the optimization models (Appendix~\ref{app:relax}). Whenever multiple methods are applied to the \emph{same} constrained search instance (namely, the true-class search or a single wrong-bit search $\mu_i$), we retain the largest valid lower bound and the smallest verified upper bound for that instance; across \emph{different} wrong-bit instances $i=1,\ldots,k$, we combine bounds via the minimum in~\eqref{eq:wrong-bound-combination}.

\subsection{Certified Cluster Decomposition}
\label{app:clusters}

A single shot often triggers several groups of detectors that are separated by many hops in the detector--mechanism Tanner graph. Solving each group independently is much faster than a joint search over all triggered detectors, provided that no candidate correction relevant to the certification verdict bridges two different groups. Topological separation in the graph alone is insufficient to guarantee this; we must prove that any correction connecting two groups exceeds a target cost threshold.

To do so, we first compute a universal lower bound $\Phi$ on the cost of any syndrome-consistent correction. Any candidate with total cost at most a chosen \emph{separation cap} $C$ then has a residual cost budget of at most $C - \Phi$ available to connect different detector groups. In our implementation, we set $C = u_0 + \delta_C$ with a margin $\delta_C > \varepsilon$ above the current true-class upper bound $u_0$. Once the groups are proved separate under the cap $C$, we solve each group independently and combine their bounds.

\subsubsection{A Universal Cost Lower Bound}

Let $\mathcal F=\{d:s_d=1\}$ be the fired detectors in the observed syndrome $s$. Every syndrome-consistent correction selects at least one mechanism that flips each detector in $\mathcal F$. To turn this observation into a lower bound, assign a non-negative charge $\varphi_d$ to each fired detector. The detector matrix entry $(\Hdec)_{dj}=1$ means that mechanism $j$ flips detector $d$, and $w_j$ is the cost~\eqref{eq:weights} of selecting that mechanism. Require that its total assigned charge does not exceed this cost:
\begin{equation}
  \sum_{d\in\mathcal F:\,(\Hdec)_{dj}=1}\varphi_d \;\le\; w_j
  \qquad\text{for every column }j.
  \label{eq:dualfeas}
\end{equation}

\begin{lemma}[Charging bound]
  \label{lem:charge}
  For any $\varphi$ satisfying \eqref{eq:dualfeas} and any $x$ with $\Hdec x=s$, $\cost(x)\ge\Phi:=\sum_{d\in\mathcal F}\varphi_d$.
\end{lemma}

\begin{proof}
  For each fired detector, choose one selected column that flips it; at least one exists because its parity is odd. Charge that detector to the chosen column. The charges assigned to any column sum to at most its weight by \eqref{eq:dualfeas}. Summing over selected columns counts every detector charge once and gives $\Phi\le\cost(x)$.
\end{proof}

We seek large charges because a larger $\Phi$ leaves less cost available to connect groups. Maximizing $\Phi$ under \eqref{eq:dualfeas} is a linear program (LP), the dual of a fractional set-cover relaxation. Optimality is not required: any charges satisfying the constraints give a valid bound. We solve the LP with HiGHS, recheck its constraints, and scale the charges down when necessary. 

\subsubsection{Cluster Separation under a Cost Cap}

The charges account for part of the weight of each mechanism. The part left over is its \emph{reduced weight},
\begin{equation}
  w'_j=w_j-\sum_{d\in\mathcal F:\,(\Hdec)_{dj}=1}\varphi_d\ge0.
  \label{eq:reducedweight}
\end{equation}
The charging argument also keeps track of these leftover weights.
Write $\mathrm{supp}(x)=\{j:x_j=1\}$ for the mechanisms selected by correction $x$. For every syndrome-consistent candidate, the total detector charge $\Phi$ and its reduced weights then give
\begin{equation}
  \cost(x)\ge\Phi+\sum_{j\in\mathrm{supp}(x)}w'_j.
  \label{eq:residualbudget}
\end{equation}
Each fired detector is flipped by at least one selected column, so expanding the weights in \eqref{eq:residualbudget} counts its charge at least once. A candidate of cost at most $C$ can therefore spend at most $C-\Phi$ in reduced weight. If $C<\Phi$, no candidate is within the cap.

To test whether groups can be connected within that budget, form a graph whose vertices are mechanism columns. Join two columns when they share any detector row, fired or not. The selected columns of a candidate form one or more connected \emph{support components}.
For two proposed groups $a$ and $b$ of fired detectors, let $\mathrm{dist}_\varphi(a,b)$ be the smallest sum of reduced weights along a path from a column incident to $a$ to one incident to $b$.
The sum includes both endpoints; it is infinite when no path exists.

\begin{proposition}[Separation]
  \label{prop:separation}
  If $\mathrm{dist}_\varphi(a,b)>C-\Phi$, no candidate of cost at most $C$ has a connected support component touching both groups.
\end{proposition}

\begin{proof}
  Such a component would contain a connecting path. Since all reduced weights are non-negative, \eqref{eq:residualbudget} would give
  \[
    \cost(x)\ge\Phi+\mathrm{dist}_\varphi(a,b)>C,
  \]
  contradicting the cap.
\end{proof}

Figure~\ref{fig:cluster_separation} illustrates the test. A connecting support component contains such a path, so its reduced weight is at least the path distance. The path itself need not satisfy the syndrome.
A short path therefore leaves separation unproved; it does not show that a connecting correction exists.

\input{fig/cluster_separation}

A labeled multi-source Dijkstra search builds the detector groups \cite{Dijkstra}. Each fired detector initially labels its incident columns.
Whenever two labels can be connected within the reduced-weight budget $C-\Phi$, they merge. Repeating these merges gives the clusters used below.

\subsubsection{Combining the Cluster Minima}

For each cluster $c$, make a smaller syndrome $s^{(c)}$ that is one on the fired detectors in $c$ and zero on every other detector.
We find its cheapest correction without fixing a logical class:
\begin{equation}
  g_c=\min\{\cost(x):\Hdec x=s^{(c)}\}.
  \label{eq:clustermin}
\end{equation}
All detector rows remain in this problem. In particular, a cluster witness must leave every detector outside its group unflipped.
If columns are filtered as in \appref{app:relax}, we solve over the retained columns; the filter preserves every global candidate within $C$.

The cluster search obtains the exact minimum $g_c$, a witness
$x^{(c)}$, and its class $\lambda_c=\Hlog x^{(c)}$.
It also checks whether another logical class can attain a cost within $\varepsilon$ of $g_c$. When it rules this out, it provides a lower bound $f_c$ for every cluster solution outside class $\lambda_c$. The bound includes unfinished frontier nodes and any candidates omitted by safe pruning.

\paragraph*{Additivity of cluster costs}
Consider a global candidate of cost at most $C$. Separation ensures that no connected component of its selected columns touches fired detectors in two different clusters. The components touching cluster $c$ together have syndrome $s^{(c)}$, so their cost is at least $g_c$.
Any components touching no fired detector have zero syndrome and non-negative cost. Thus every candidate within the cap costs at least $G:=\sum_c g_c$.

Conversely, XORing the cluster witnesses produces a correction with syndrome $s$ and cost at most $G$: overlapping selected mechanisms cancel. If $G\le C$, the lower bound just proved forces this correction to cost exactly $G$, so it is a global minimum. Positive weights then imply that the witnesses have disjoint supports. Its logical class is $\lambda_*=\bigoplus_c\lambda_c$. We verify the syndrome, cost, and class of the composed witness before using it.

\paragraph*{Lower bounds on alternative logical classes} Knowing a cheapest correction is not enough for a verdict; we also need to bound every competing logical class. Within the cap, a candidate outside $\lambda_*$ must use a different class for at least one cluster or include a nonempty zero-syndrome component. Changing cluster $c$ costs at least $f_c-g_c$ more than its minimum. A nonempty zero-syndrome component costs at least $w_{\min}$, the smallest mechanism weight in the model. Hence
\begin{equation}
  \begin{aligned}
    \cost(x)&\ge G+D_{\min}\quad
    (\Hlog x\ne\lambda_*,\ \cost(x)\le C),\\
    D_{\min}&=\min\left\{\min_c(f_c-g_c),\;w_{\min}\right\}.
  \end{aligned}
  \label{eq:composition}
\end{equation}

Candidates outside the cap cost more than $C$. Combining them with the within-cap bound gives a lower bound on every class other than
$\lambda_*$:
\begin{equation}
  L_{\mathrm{alt}}=\min\{C,\;G+D_{\min}\}
  \label{eq:clusteralt}
\end{equation}
If $L_{\mathrm{alt}}>G+\varepsilon$, the composed witness beats every other class by the required margin. The sector succeeds when $\lambda_*=\lambda_0$ and fails otherwise. The cap must therefore exceed $G+\varepsilon$ for this test to reach a verdict.
Figure~\ref{fig:cluster_compose} summarizes these bounds for two clusters.

\input{fig/cluster_compose}

If the cluster search finds a near-tie, reaches a resource limit, or cannot prove a sufficient gap, the sector proceeds to the searches of Section~\ref{sec:engine} and \appref{app:relax}.

\subsection{Optimization Bounds and Integer Programming}
\label{app:relax}

Optimization models help when the search bounds still overlap. In this section, $Ax=b$ over $\mathbb F_2$ denotes one active problem. The matrix $A$ contains all detector rows and whichever logical rows the true-class or wrong-bit search requires; $b$ contains their target parity bits. The correction vector $x$ has one entry for each of the $n$ mechanism columns, and $w_j$ is the positive cost of mechanism $j$.
Every model minimizes the total cost $w^{\mathsf T}x$.

We use models in two ways. A \emph{relaxation} contains every binary correction and possibly fractional points, so its optimum bounds the binary minimum from below. A verified binary correction bounds it from above. Matching bounds establish the minimum. The final subsection states what numerical verification these claims rely on.

\subsubsection{Linear Programming Relaxations}
\label{sec:relax-main}

The simplest relaxation allows $0\le x_j\le1$ and keeps only inequalities satisfied by every binary solution. We strengthen it with an adaptive parity-cut linear program (LP) \cite{feldman2005lp,taghavi2008adaptive}. Its added inequalities rule out some fractional points while retaining all binary corrections. Every completed LP solve therefore gives a lower bound, even if the cut-generation budget ends early. An LP solution gives an upper bound only when it is binary and passes the detector and logical checks.

\subsubsection{Encoding Parity with Integer Variables}

Before adding stronger parity constraints, we express each parity row as an integer equation. For row $r$ of the active matrix $A$, let $S_r=\{j:A_{rj}=1\}$ be the mechanism columns participating in that row, and let $b_r\in\{0,1\}$ be its target parity. The binary variable $x_j$ selects mechanism $j$; introduce an auxiliary variable $z_r$ to write
\begin{equation}
  \sum_{j\in S_r}x_j-2z_r=b_r,
  \qquad x_j\in\{0,1\},\quad z_r\in\mathbb{Z}_{\ge0}.
  \label{eq:parityinteger}
\end{equation}
Here $z_r$ counts pairs of selected columns beyond the required parity bit. For example, selecting three columns in an odd row gives $3-2z_r=1$, hence $z_r=1$. Each binary solution fixes exactly one $z_r=(\sum_{j\in S_r}x_j-b_r)/2$, so the equation encodes parity exactly.

If both $x$ and $z$ are made continuous, the same equation requires only $\sum_{j\in S_r}x_j\ge b_r$. An even row adds no restriction beyond $0\le x_j\le1$; an odd row requires only total mass at least one. Thus the continuous model loses much of the parity information.

\subsubsection{Adaptive Parity-Cut Linear Program}
\label{sec:parity-cut-lp}

The parity-cut LP keeps $x\in[0,1]^n$ and adds inequalities satisfied by every valid binary assignment. For a row, choose $T\subseteq S_r$ with the \emph{wrong} parity, $|T|\not\equiv b_r\pmod2$. The assignment that is one exactly on $T$ violates that row.
Every valid binary assignment must differ in at least one position, so
\[
  \sum_{j\in T}(1-x_j)+\sum_{j\in S_r\setminus T}x_j\ge1.
\]
Rearranging gives the Feldman forbidden-set inequality \cite{feldman2005lp}
\begin{equation}
  \sum_{j\in T}x_j-\sum_{j\in S_r\setminus T}x_j\le |T|-1.
  \label{eq:forbidden}
\end{equation}

There can be exponentially many cuts per row, so we add them only as needed \cite{taghavi2008adaptive}. After solving the current LP, form $T$ from entries with $x_j>1/2$. If $T$ has the required parity, reverse membership of the entry closest to $1/2$. This chooses a most violated forbidden-set inequality for that row. We add it when its violation exceeds the tolerance and solve again. Before insertion, we check that $T$ has the wrong parity; otherwise the cut could exclude a valid correction and make the lower bound unsound.

If the LP optimum is binary and satisfies the active rows, it is a witness attaining the binary minimum. Otherwise, its objective remains only a lower bound.

\subsubsection{Integer Programming and Cut Reuse}
\label{sec:ip}

An intermediate model makes $z$ integral in
\eqref{eq:parityinteger} but leaves $x\in[0,1]^n$. Each row sum then
has the required integer parity. Fractional $x$ can still remain, so
the optimum is a lower bound, not necessarily a correction. This
small model can show that no candidate lies within the cost cap
without solving the full binary problem.

If the sector remains undecided, we enforce binary constraints $x \in \{0,1\}^n$ and solve the resulting mixed-integer program (MIP) with HiGHS~\cite{huangfu2018highs} subject to a time limit. We set the solver's relative optimality gap tolerance to zero so that the search does not terminate early before the dual lower bound matches the best integer objective. Throughout the solve, valid lower bounds are taken exclusively from the solver's dual bound, whereas upper bounds are taken from the \emph{incumbent} i.e., the cheapest feasible binary solution found so far, after independently verifying its parities. We treat a class minimum as exact only if the dual bound and the verified incumbent cost agree to within $10^{-10}$ (well below $\varepsilon$); if the solver hits the time limit first, both bounds remain valid and are retained.

\subsubsection{Integer Feasibility Tests}
\label{sec:feasibility}

A sector can be decided without finding an exact minimum. Let $\hat x$ be a verified correction with cost $c$ and class $\hat\lambda$, and choose a margin $\delta>\varepsilon$, where $\varepsilon$ is the verdict tolerance.

When $\hat\lambda=\lambda_0$, test whether any competing correction costs at most $c+\delta$. If none exists, then the true-class upper bound satisfies $u_0\le c$ and the competing-class lower bound satisfies $\ell_w\ge c+\delta>u_0+\varepsilon$. Proposition~\ref{prop:verdict}
therefore certifies success.

When $\hat\lambda\ne\lambda_0$, apply the same cost cap to the true class.
If no true-class correction lies within the cap, then the competing-class upper bound satisfies $u_w\le c$ and the true-class lower bound satisfies $\ell_0\ge c+\delta>u_w+\varepsilon$, which certifies failure.

Both conditions are integer feasibility problems with a zero objective. For the true class, we encode the detector constraints $\Hdec x=s$ and the logical constraints $\Hlog x=\lambda_0$ using \eqref{eq:parityinteger}. We add the cost constraint $w^{\mathsf T}x\le c+\delta$.

For the competitors, we leave the logical class free and record its $k$ bits in the binary vector $y\in\{0,1\}^k$. The equality $\Hlog x-2q-y=0$ encodes these logical parities, with each nonnegative integer $q_i$ counting pairs of selected mechanisms in logical row $i$.
A single linear inequality then requires $y$ to differ from the true class $\lambda_0$ in at least one bit:
\begin{equation}
  \sum_{i:\lambda_{0,i}=0}y_i+\sum_{i:\lambda_{0,i}=1}(1-y_i)\ge1.
  \label{eq:anywrong}
\end{equation}
One problem thus covers all $2^k-1$ competing classes, as the union in Lemma~\ref{lem:union} does. An infeasibility proof establishes the required bound. A feasible point is a correction within $c+\delta$, which we verify before using it as a witness.

We obtain $\hat x$ from the syndrome-only integer program, solved as above, and try margins $\delta$ from $0.1$ down to $10^{-6}$ nats. A larger margin establishes a wider cost gap when the test proves infeasibility; smaller margins allow verdicts for closer minima. The verdict does not depend on $\hat x$ being optimal: its class only selects which test to run, and its cost only sets the cap. Consequently, this route relies on the infeasibility claims of the solver but on no optimality claim.

\subsubsection{Reusing Charges to Prune the Search}
\label{sec:charge-reuse}

The detector charges of \appref{app:clusters} are computed once per shot and can also help individual searches. At a search node, some originally fired detectors still require a flip from the unassigned columns. Their charges bound the additional cost needed to complete the node. We update this charge bound as rows change parity and take the stronger of it and the disjoint-row bound.

The same charges identify columns that cannot occur in a cheap candidate. By \eqref{eq:residualbudget}, any candidate containing column $j$ costs at least $\Phi+w'_j$. If this exceeds the working cap, we can remove $j$ from the search and optimization models when considering candidates within that cap. Numerical slack is allowed before removal. The test applies to every logical target because it uses only the detector syndrome.

The restricted models still contain every candidate within the cap. To use one of their bounds for the unrestricted problem, we also account for excluded candidates, all of which cost more than the cap, as in \eqref{eq:clusteralt}. At low physical error rates, $\Phi$ is close to the cap, so few columns have small enough reduced weight to remain. On identical shots at $p=6\times10^{-3}$, this filter reduced median search expansions by roughly two orders of magnitude and wall time by a factor of six, without contradictory verdicts. No integer solves were needed at the lowest sampled rates.

\subsubsection{Numerical Verification}

Every binary witness is checked independently against the detector and logical equations using integer arithmetic modulo two. Its computed cost is then an upper bound. This check establishes feasibility, not optimality or the validity of the lower bound reported by a solver.

Costs, charges, and solver bounds use floating-point arithmetic. The current results therefore rely on the solver and its tolerances for numerical lower bounds and for the infeasibility claims of \appref{app:relax}. The margin $\varepsilon$ avoids verdicts near a tie, but does not bound every rounding error.

We also guard two presolve operations, which simplify the optimization model before the main solve. Bounds on auxiliary integer variables are rounded down to preserve their allowed integer values. Each parity-cut row receives a finite lower bound equal to the sum of its negative coefficients, which is valid for $0\le x_j\le1$. Neither change removes a feasible binary correction.

Exhaustive tests on small instances and independent solver comparisons provide additional validation. In particular, for the $[[144,12,12]]$ code across all sampled physical error rates, we re-evaluated every certified failure using the exact integer-programming decoder bundled with Tesseract~\cite{aghababaie2025tesseract} (whose outputs were never used to seed the certification search). On every tested shot, the independent solver confirmed a competing logical class.

\fi

\end{document}

%% file: fig/styles.tex
%
\tikzset{
  cert/.style={font=\sffamily\scriptsize, >={Latex[length=4pt,width=3pt]},
    line join=round},
  panel/.style={font=\sffamily\scriptsize\bfseries, anchor=base west,
    inner sep=0pt},
  muted/.style={text=certGray!90!black},
  even/.style={circle, inner sep=0pt, minimum size=4.4pt, fill=certBlue,
    draw=certBlue},
  odd/.style={circle, inner sep=0pt, minimum size=4.4pt, fill=white,
    draw=certOrange!90!black, line width=0.7pt},
}

%% file: fig/roadmap_full.tex
\begin{figure}[!t]
  \centering
  \begin{tikzpicture}[cert,
      stage/.style={draw=certGray!80, fill=white, rounded corners=2pt,
        line width=0.6pt, inner sep=3pt, align=left, anchor=north west,
      text width=49mm},
      accel/.style={stage, draw=certBlue!80, fill=certBlueLight,
      text width=45.5mm},
      outcome/.style={rounded corners=2pt, line width=0.7pt, inner sep=3pt,
      align=center, text width=21mm},
      flow/.style={->, line width=0.6pt, draw=certGray},
    feed/.style={line width=0.6pt, draw=certOrange!90!black, densely dashed}]
    \coordinate (flowx) at (3.1,0);   
    \coordinate (skipx) at (0.625,0); 
    \coordinate (costx) at (0.15,0);  
    \coordinate (bus)   at (5.82,0);  

    \node[stage, fill=certGrayLight] (s1) at (0.45,0) {%
      \textbf{1.\ Sample and initialize}\hfill
      {\color{certGray}Sec.~\ref{sec:setup}--\ref{sec:certs}}\\
      The sampled error gives $s$, $\lambda_0$, and $u_0$; the detector
    bound~\eqref{eq:detector-bound} gives $\ell_0,\ell_w$.};
    \node[accel] (s2) at ($(s1.south west)+(0.35,-0.25)$) {%
      \textbf{2.\ Seed and reduce}\hfill
      {\color{certGray}App.~\ref{sec:anchor}--\ref{sec:unified}}\\
      Witnesses lower $u_0,u_w$; $k{+}1$ wrong-bit problems or one
    syndrome-only search.};
    \node[accel] (s3) at ($(s2.south west)+(0,-0.25)$) {%
      \textbf{3.\ Charges and clusters}\hfill
      {\color{certGray}App.~\ref{app:clusters}, \ref{sec:charge-reuse}}\\
      Charges $\Phi$ filter columns; separated clusters give a witness and
    $L_{\mathrm{alt}}$.};
    \node[accel] (s4) at ($(s3.south west)+(0,-0.25)$) {%
      \textbf{4.\ Parity-cut LP}\hfill
      {\color{certGray}App.~\ref{sec:relax-main}--\ref{sec:parity-cut-lp}}\\
      Forbidden-set cuts raise $\ell_0,\ell_w$; a binary LP optimum is a
    witness.};
    \node[stage] (s5) at ($(s4.south west)+(-0.35,-0.25)$) {%
      \textbf{5.\ Branch and bound}\hfill
      {\color{certGray}Sec.~\ref{sec:engine}}\\
      Best-first search; frontier and retained bounds stay valid at resource
    limits.};
    \node[accel] (s6) at ($(s5.south west)+(0.35,-0.25)$) {%
      \textbf{6.\ Integer programs}\hfill
      {\color{certGray}App.~\ref{sec:ip}--\ref{sec:feasibility}}\\
    MIP dual bounds and zero-objective feasibility tests.};

    \foreach \a/\b in {s1/s2, s2/s3, s3/s4, s4/s5, s5/s6}
    \draw[flow] (\a.south -| flowx) -- (\b.north -| flowx);
    \draw[flow, densely dotted] (s1.south -| skipx) -- (s5.north -| skipx);
    \draw[->, line width=0.8pt, certGray!55] (s1.north -| costx)
    -- (s6.south -| costx)
    node[midway, rotate=90, anchor=south, inner sep=2pt, text=certGray]
    {increasing computational cost};

    \foreach \s in {s1,s2,s3,s4,s5,s6}
    \draw[feed] (\s.east) -- (\s.east -| bus);
    \draw[feed] (s1.east -| bus) -- (s6.east -| bus);
    \node[outcome, draw=certOrange!90!black, fill=certOrangeLight,
    anchor=west] (cmp) at ($(s1.north -| bus)!0.5!(s6.south -| bus)
    + (0.23,0)$) {%
      \textbf{Compare intervals}\\(Prop.~\ref{prop:verdict})\\[2pt]
      success if\\$\ell_w>u_0+\varepsilon$\\[2pt]
    failure if\\$u_w<\ell_0-\varepsilon$};
    \draw[feed, ->] (cmp.west -| bus) -- (cmp.west);
    \node[outcome, draw=certGreen!80!black, fill=certGreenLight,
    above=6mm of cmp] (done) {%
    \textbf{Certified}\\success or failure;\\the sector stops};
    \node[outcome, draw=certRed!80, fill=certRedLight,
    below=6mm of cmp] (open) {%
    \textbf{Undecided}\\tie or budget limit;\\counted in $[\LB,\UB]$};
    \draw[->, line width=0.8pt, certGreen!80!black] (cmp) -- (done)
    node[midway, right, text=certGreen!80!black] {separated};
    \draw[->, line width=0.8pt, certRed!80] (cmp) -- (open)
    node[midway, right, text=certRed!80!black] {after step 6};
  \end{tikzpicture}
  \caption{Certification workflow for one sector. Shaded, indented steps are
    optional accelerations; the dotted arrow is the baseline procedure of
    Sections~\mbox{\ref{sec:setup}--\ref{sec:engine}}. Each step may lower an upper
    bound ($u_0$, $u_w$) with a verified witness or raise a lower bound
    ($\ell_0$, $\ell_w$). The intervals are compared after every update
    (Proposition~\ref{prop:verdict}), so a sector stops as soon as it is
    certified. A near-tie, or a sector still open after the last step, is
  recorded as undecided.}
  \label{fig:roadmap-full}
\end{figure}
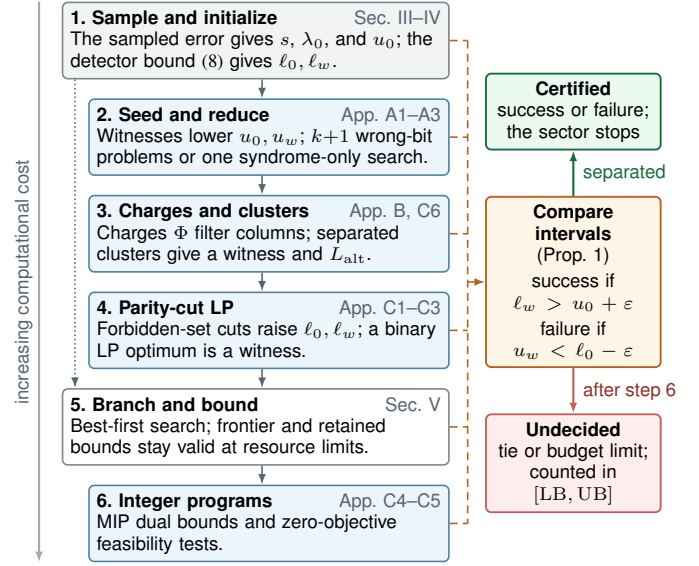

%% file: fig/cluster_separation.tex
\begin{figure}[!t]
  \centering
  \begin{tikzpicture}[cert,
      det/.style={rectangle, minimum size=5pt, inner sep=0pt,
      line width=0.6pt},
      fired/.style={det, fill=certGray!85!black, draw=certGray!85!black},
      unfired/.style={det, fill=white, draw=certGray},
      group/.style={draw=certBlue!80, fill=certBlueLight, rounded corners=3pt,
      line width=0.6pt},
      charge/.style={fill=certBlue!30, draw=certBlue!80!black,
      line width=0.5pt},
      reduced/.style={fill=certOrange!30, draw=certOrange!90!black,
    line width=0.5pt}]
    \node[panel] at (0,2.3) {(a) Cheapest path from group $a$ to group $b$};
    \draw[group] (0.05,-0.12) rectangle (1.0,0.42);
    \node[anchor=south, text=certBlue!90!black] at (0.525,0.42) {group $a$};
    \draw[group] (4.85,-0.12) rectangle (5.8,0.42);
    \node[anchor=south, text=certBlue!90!black] at (5.325,0.42) {group $b$};
    \draw[line width=0.7pt, certGray] (0.7,0.15) -- (5.15,0.15);
    \draw[charge] (1.35,0.15) rectangle (1.85,0.95);
    \draw[reduced] (1.35,0.95) rectangle (1.85,1.19);
    \draw[reduced] (2.75,0.15) rectangle (3.25,0.71);
    \draw[charge] (4.15,0.15) rectangle (4.65,0.95);
    \draw[reduced] (4.15,0.95) rectangle (4.65,1.55);
    \foreach \x in {0.35,0.7,5.15,5.5} \node[fired] at (\x,0.15) {};
    \foreach \x in {2.3,3.7} \node[unfired] at (\x,0.15) {};
    \node[text=certBlue!90!black] at (1.6,0.55) {2.0};
    \node[text=certBlue!90!black] at (4.4,0.55) {2.0};
    \node[anchor=south] at (1.6,1.19) {$w'_{j_1}{=}0.6$};
    \node[anchor=south] at (3.0,0.71) {$w'_{j_2}{=}1.4$};
    \node[anchor=south] at (4.4,1.55) {$w'_{j_3}{=}1.5$};
    \foreach \x/\j in {1.6/1,3.0/2,4.4/3} \node[anchor=north] at (\x,0.1) {$j_\j$};
    \draw[charge] (6.25,1.3) rectangle (6.5,1.45);
    \node[anchor=west] at (6.55,1.375) {detector charges};
    \draw[reduced] (6.25,0.98) rectangle (6.5,1.13);
    \node[anchor=west] at (6.55,1.055) {reduced weight $w'_j$};
    \node[fired] at (6.375,0.73) {};
    \node[anchor=west] at (6.55,0.73) {fired detector};
    \node[unfired] at (6.375,0.43) {};
    \node[anchor=west] at (6.55,0.43) {unfired detector};

    \node[panel] at (0,-0.5) {(b) Cost of any candidate connecting $a$ and $b$};
    \draw[charge] (0.3,-1.5) rectangle (4.5,-1.1);
    \node[text=certBlue!90!black] at (2.4,-1.3) {charges $\Phi=7$};
    \draw[reduced] (4.5,-1.5) rectangle (4.86,-1.1);
    \draw[reduced] (4.86,-1.5) rectangle (5.7,-1.1);
    \draw[reduced] (5.7,-1.5) rectangle (6.6,-1.1);
    \fill[pattern=north east lines, pattern color=certRed] (6.3,-1.5)
    rectangle (6.6,-1.1);
    \draw[decorate, decoration={brace, amplitude=3pt}, certGray]
    (4.5,-1.02) -- (6.26,-1.02) node[midway, above=2pt, muted]
    {$C-\Phi=3$};
    \draw[decorate, decoration={brace, amplitude=3pt, mirror},
    certOrange!90!black] (4.5,-1.58) -- (6.6,-1.58)
    node[midway, below=2pt, text=certOrange!90!black]
    {$\mathrm{dist}_\varphi(a,b)=3.5$};
    \draw[certRed, line width=1pt] (6.3,-1.62) -- (6.3,-0.98)
    node[anchor=south west, inner sep=1pt, text=certRed] {$C=10$};
    \node[anchor=west, text=certRed] at (6.7,-1.3) {cost $\ge10.5>C$};
  \end{tikzpicture}
  \caption{Separation by reduced cost, with illustrative values. (a)~Columns
    $j_1$, $j_2$, $j_3$ form the cheapest path between groups $a$ and $b$ of
    fired detectors; consecutive columns share a detector. Each weight $w_j$
    splits into the charges of the fired detectors that column~$j$ flips and
    its reduced weight $w'_j$~\eqref{eq:reducedweight}. (b)~Every
    syndrome-consistent candidate pays the total charge $\Phi=7$. A candidate
    whose support connects $a$ and $b$ also pays at least
    $\mathrm{dist}_\varphi(a,b)=0.6+1.4+1.5=3.5$ in reduced
    weight~\eqref{eq:residualbudget}. Since $3.5>C-\Phi=3$, it costs more
  than the cap $C=10$ (Proposition~\ref{prop:separation}).}
  \label{fig:cluster_separation}
\end{figure}

%% file: fig/cluster_compose.tex
\begin{figure}[!t]
  \centering
  \begin{tikzpicture}[cert,
      gone/.style={fill=certBlue!30, draw=certBlue!80!black, line width=0.5pt},
      gtwo/.style={fill=certBlue!15, draw=certBlue!80!black, line width=0.5pt},
      extra/.style={fill=certOrange!30, draw=certOrange!90!black,
        line width=0.5pt},
      refline/.style={line width=0.6pt},
      lab/.style={anchor=north, align=center, inner sep=1pt}]
    \draw[refline, certBlue, densely dashed] (0.95,1.5) -- (8.05,1.5);
    \draw[refline, certGreen!80!black, densely dotted] (0.95,1.8) -- (8.05,1.8);
    \draw[refline, certOrange!90!black, densely dashed] (0.95,2.25) -- (8.05,2.25);
    \draw[refline, certRed] (0.95,3.0) -- (8.05,3.0);
    \node[anchor=east, text=certBlue] at (0.9,1.5) {$G$};
    \node[anchor=east, text=certGreen!70!black] at (0.9,1.8) {$G{+}\varepsilon$};
    \node[anchor=east, text=certOrange!90!black] at (0.9,2.25)
      {$L_{\mathrm{alt}}$};
    \node[anchor=east, text=certRed] at (0.9,3.0) {cap $C$};
    \draw[->, line width=0.7pt, certGray] (1.0,0) -- (1.0,3.5)
      node[above, muted] {cost};
    \foreach \x/\h/\name in {1.3/1.5/, 2.7/2.25/{$f_1{-}g_1$\\${=}D_{\min}$},
        4.1/2.65/{$f_2{-}g_2$}, 5.5/2.45/{$w_{\min}$}} {
      \draw[gone] (\x,0) rectangle (\x+0.9,0.85);
      \draw[gtwo] (\x,0.85) rectangle (\x+0.9,1.5);
      \node at (\x+0.45,0.425) {$g_1$};
      \node at (\x+0.45,1.175) {$g_2$};
      \ifdim\h pt>1.5pt
        \draw[extra] (\x,1.5) rectangle (\x+0.9,\h);
        \node[align=center, inner sep=0pt] at (\x+0.45,{(1.5+\h)/2})
          {\name};
      \fi
    }
    \fill[certRedLight] (6.9,0) rectangle (7.8,3.0);
    \fill[pattern=north east lines, pattern color=certRed!70] (6.9,3.0)
      rectangle (7.8,3.35);
    \draw[certRed!80, line width=0.5pt] (6.9,3.35) -- (6.9,0) -- (7.8,0)
      -- (7.8,3.35);
    \node[text=certRed!80!black] at (7.35,2.62) {$>C$};
    \node[lab] at (1.75,-0.05) {composed\\witness $\lambda_*$};
    \node[lab] at (3.15,-0.05) {cluster 1 in\\other class};
    \node[lab] at (4.55,-0.05) {cluster 2 in\\other class};
    \node[lab] at (5.95,-0.05) {zero-\\syndrome\\component};
    \node[lab] at (7.35,-0.05) {outside\\the cap};
  \end{tikzpicture}
  \caption{Bounding the other classes after composing two separated
    clusters, \eqref{eq:composition}--\eqref{eq:clusteralt}. The composed
    witness costs $G=g_1+g_2$ and has class $\lambda_*=\lambda_1\oplus
    \lambda_2$. Within the cap, a correction in another class must change
    the class of some cluster~$c$, adding at least $f_c-g_c$, or contain a
    nonempty zero-syndrome component, adding at least $w_{\min}$; any other
    correction costs more than $C$. Every other class therefore costs at
    least $L_{\mathrm{alt}}=\min\{C,G+D_{\min}\}$, and
    $L_{\mathrm{alt}}>G+\varepsilon$ decides the sector.}
  \label{fig:cluster_compose}
\end{figure}
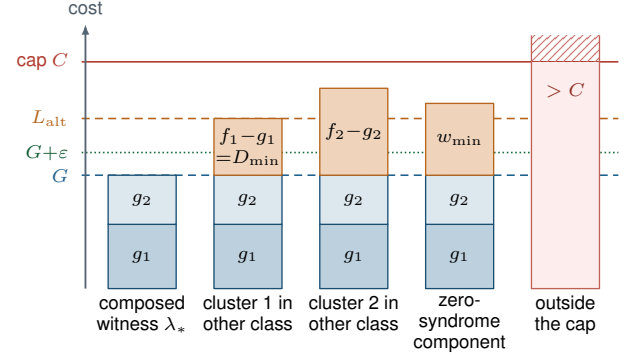